\documentclass[11pt]{article}
\usepackage{a4wide}
\usepackage[dvips]{graphicx}
\usepackage[francais]{babel}
\usepackage{times}
\usepackage[latin1]{inputenc}
\usepackage{eurosym}
\def\LASEP{
{\rm L
\kern-0.36em\raise.3ex\hbox{\sc a}
\kern-.15emS
\kern-.1667em\lower.7ex\hbox{E}
\kern-.125emP}}
\begin{document}
\pagestyle{empty} \noindent
\begin{center}
\large{\textbf{Calibration d'une lampe à filament de tungstène}}
\end{center}
\rule[0pt]{\textwidth}{0.1pt}\\
\rule[0pt]{0pt}{0.5cm}\\
\noindent
\emph{\textbf{Charles de IZARRA}}\\

\noindent \emph{Groupe de Recherche sur l'\'Energetique des
Milieux Ionisés,\\ UMR6606 Université d'Orléans - CNRS, Facult\'e
des Sciences, Site de Bourges,\\ rue Gaston Berger, BP
4043, 18028 BOURGES Cedex, France.\\
Charles.De$\_$Izarra@univ-orleans.fr\\
Tél : 33 (0)2 48 27 27 31}\\
\rule[0pt]{\textwidth}{0.1pt}\\
\rule[0pt]{0pt}{0.5cm}\\
\noindent \textbf{Résumé}\\
\noindent \emph{L'objet du travail présenté est la calibration
d'une lampe à filament de tungstène à partir de mesures
électriques à la fois simples et précises, et qui permettent de
déterminer la température du filament de tungstène en fonction de
l'intensité du courant d'alimentation, puis de mesurer la surface
du filament. Ces données permettent alors de calibrer la lampe en
terme de luminance, directement utilisable pour étalonner des
capteurs comme des pyromètres optiques ou tout autre capteur
photométrique. La procédure de calibration proposée a été
appliquée sur une lampe à filament standard (lampe utilisée dans
l'éclairage automobile), puis sur une lampe étalon à ruban de
tungstène. La procédure de calibration mise au point permet de
retrouver les points de calibration de la lampe (NIST) avec une
précision de l'ordre de 2\%.
 }\\ \ \\
\noindent \emph{Mots-Clés : Lampe à ruban de tungstène, Photométrie, Rayonnement thermique}\\
\rule[0pt]{\textwidth}{0.1pt}\\
\rule[0pt]{0pt}{0.5cm}\\

\noindent\textbf{Introduction}\\
Les capteurs photométriques tels que les pyromètres optiques,
destinés à mesurer la température de corps chauffés nécessitent
d'être calibrés en enregistrant le signal qu'ils délivrent en
analysant un corps chauffé dont on connaît à la fois la
température et l'émissivité. Le corps de référence choisi est
souvent un corps noir de laboratoire, dont le principal
inconvénient est d'être limité à des températures assez basses
(inférieures à 2000 K pour les plus courants). Une autre
possibilité est l'utilisation d'une lampe étalon (mère ou fille) à
ruban de tungstène, dont le prix est relativement élevé (de
l'ordre de 7000 \euro), dont on connaît la température pour une
valeur de l'intensité du courant continu utilisé pour alimenter la
lampe. Dans cette communication, on expose une technique de
calibration d'une lampe à filament de tungstène avec des mesures
électriques alliant simplicité et précision, et avec lesquelles on
détermine la température du filament de tungstène en fonction de
l'intensité du courant d'alimentation.\\

\noindent\textbf{1. L'émission radiative des corps opaques chauffés}\\
La surface d'un corps que nous choisirons opaque, à la température
$T$, est la source d'une émission de rayonnement continu dont la
répartition de l'énergie en fonction de la longueur d'onde est
donnée par la loi de Planck [Chéron, 1999], basée sur le modèle du
corps noir, qui, par définition est capable d'absorber toutes les
longueurs d'onde $\lambda$ qu'il reçoit.  Quantitativement, on
utilise la luminance spectrale $L_\lambda ^0$ donnée par :
\begin{equation}
L_\lambda ^0 = \frac{2 h c ^2}{\lambda ^5}
\frac{1}{\exp\left(\frac{hc}{\lambda kT}\right) -1} \label{Planck}
\end{equation}
avec :
\begin{itemize}
\item [$\bullet$] $h$ : constante de Planck ($h$ = 6.6 10 $^{-34}$ J.s)
\item [$\bullet$] $k$ : constante de Boltzmann ($k$ = 1.38 10 $^{-23}$ J/K)
\item [$\bullet$] $c$ : célérité de la lumière dans le vide ($c$ = 3 10 $^{8}$ m/s)
\end{itemize}
et qui  est le flux émis par unité de surface apparente, par unité
d'angle solide et par unité de longueur d'onde (Unité : W m$^{-2}$
Sr$^{-1}$ m$^{-1}$).

Dans le cas où le corps émissif est une surface $S$ de corps noir,
la puissance émise dans le demi-espace est donnée par la relation
de Stefan-Boltzmann :
\begin{equation}
P^0 = S \sigma T^4 \label{Stefan}
\end{equation}
où $\sigma$ est la constante de Stephan ($\sigma = 5.67 10 ^{-8}$
W.K$^{-4}$.m$^{-2}$). Lorsque le corps considéré n'est pas un
corps noir, on parle alors de corps gris ou de corps réel, et les
données précédentes doivent être réduites en les multipliant par
l'émissivité $\varepsilon(\lambda, T)$ inférieure à 1.\\

\noindent\textbf{2. Les données thermophysiques du tungstène}\\
Le tungstène est le métal possédant la température du point de
fusion la plus élevée ($T_F$=3695~K). De ce fait, il a été
largement étudié depuis le début du XX$^{\grave{e}me}$ siècle afin
de permettre la production de filaments de lampes à incandescence
[Forsythe and Worthing, 1916].

Sachant que le filament de la lampe est essentiellement une
résistance morte chauffée par effet Joule, les données nécessaires
au calcul de la résistance en fonction de la température sont la
résistivité $\rho$ et le coefficient de dilatation thermique
$\beta$. Ces deux grandeurs sont tabulées en fonction de la
température, et $\beta$ est donné en \% de la longueur $\ell_0$ du
filament à la température de 300 K [Lide, 1991-1992].

Pour une température $T$ donnée, la résistance $R$ d'un fil de
tungstène est donnée par la relation : $R(T)=\rho(T)
{\ell(T)}/{S_f(T)}$ où $S_f$ est la section du fil, que sous
supposons constante sur sa longueur. La section, pas
nécessairement circulaire, peut être exprimée en fonction d'une
longueur caractéristique $d$. Nous avons $S=C d^2$, où $C$ est une
constante de proportionnalité qui dépend de la forme géométrique
de la section. Pour fixer les idées, $C=\pi/4$ dans le cas d'une
section circulaire en choisissant le diamètre du fil comme
longueur caractéristique $d$.

Calculons la variation de la résistance $R(T)$ relativement à la
résistance $R_0$ à la température de 300~K prise comme référence.
Nous avons : $R_0=\rho_0 {\ell_0}/{{S_f}_0}$ ou encore, en
introduisant la longueur $d_0$ à  300~K : $R_0= {\rho_0 \ell_0}/{C
d_0 ^2}.$ Pour une température $T$, la résistance $R(T)$ est :
$$R(T)= \frac{\rho(T) \ell(T)}{C d^2(T)} \Rightarrow
R(T)= \frac{\rho(T) \left[ \ell_0 + \frac{\beta \ell_0}{100}
\right]}{\left[ d_0 + \frac{\beta d_0}{100} \right]^2} .$$

Le rapport $R(T)/R_0$ est alors :
\begin{equation}
R(T)/R_0 = \frac{100 \rho(T)}{\left[100+\beta\right] \rho_0}.
\label{rapport}
\end{equation}

\begin{figure}[h!]
\centerline{\includegraphics[scale=0.9,
angle=-90]{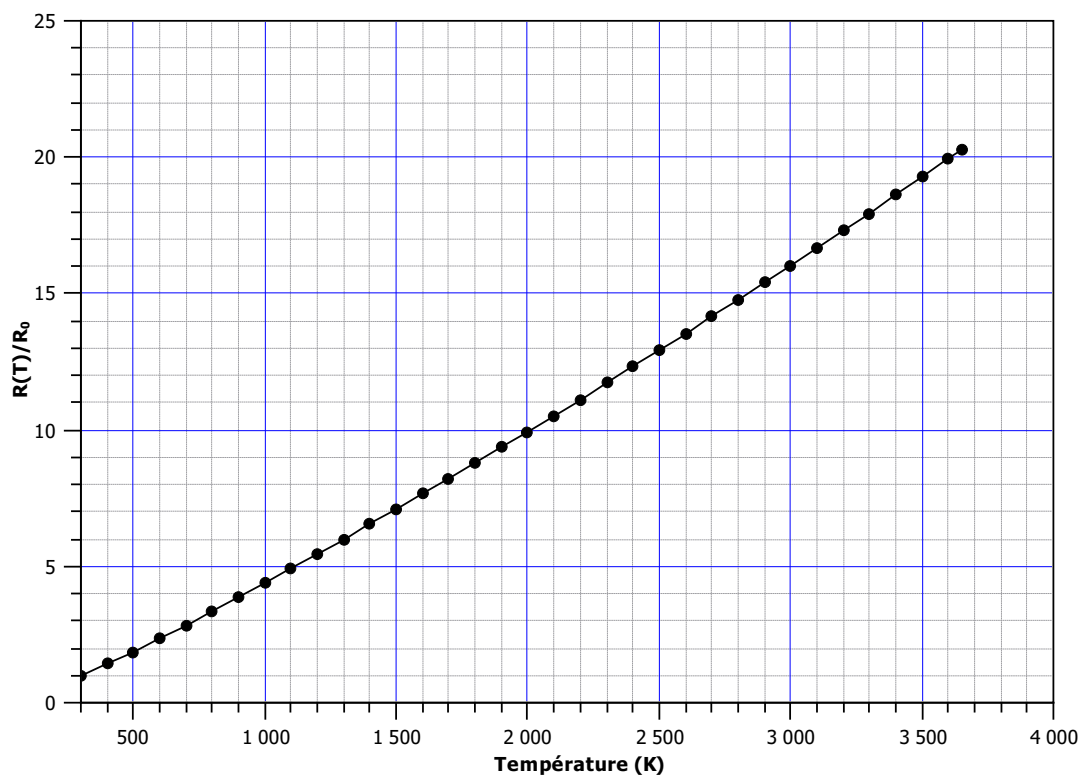}} \caption{\textit{Courbe
représentative de $R(T)/R_0$ en fonction de la température $T$
pour une résistance de tungstène.} } \label{COURBE1}
\end{figure}

La relation (\ref{rapport}), permet, à partir des données
disponibles dans la littérature, de calculer le rapport $R(T)/R_0$
en fonction de la température $T$. La courbe représentative de
$R(T)/R_0$ en fonction de $T$ est donnée sur la figure
\ref{COURBE1}; on remarque que lorsque la température augmente de
3000~K, la résistance est multipliée par 20.

Nous avons choisi de déterminer par une méthode des moindres
carrés l'équation d'une parabole passant par les couples de points
($R(T)/R_0,T)$ avec une erreur relative inférieure à 0.1\% pour
les températures élevées (voir équation (\ref{FIT})).

\begin{equation}
R(T)/R_0=-0,52427113+0,00466128 T +2,8420718 10^{-7}T^2
\label{FIT}
\end{equation}

\begin{figure}[h!]
\centerline{\includegraphics[scale=.7]{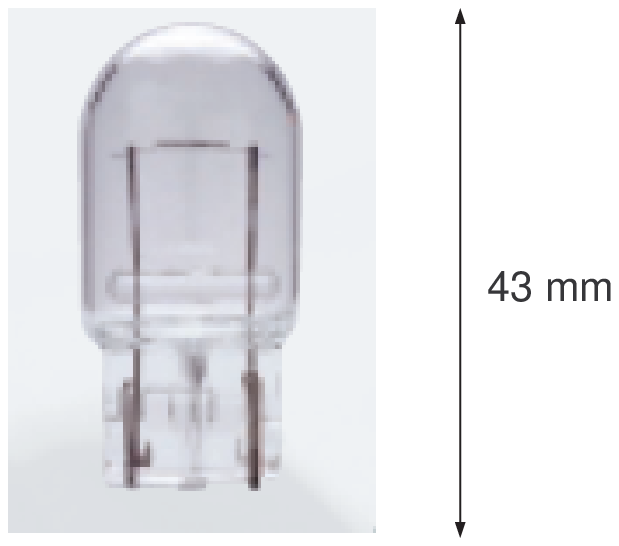}\includegraphics[scale=.7]{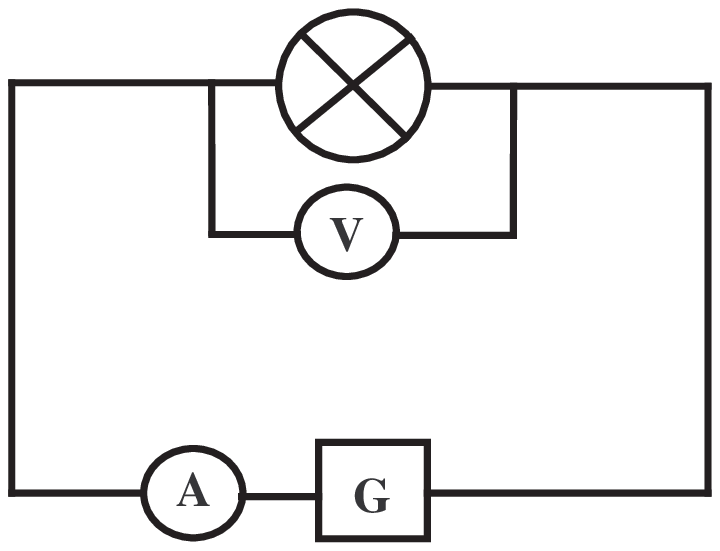}}
\caption{\textit{Lampe à filament de tungstène utilisée (image de
gauche) et montage électrique utilisé (image de droite).} }
\label{MONTAGE}
\end{figure}

\noindent\textbf{3. Procédure de calibration d'une lampe à filament de tungstène}\\
Les mesures ont été réalisées en utilisant une lampe Philips de
type E4-2DT W21W, prévue pour fonctionner dans des conditions
nominales sous une tension continue de 12 V, pour une puissance de
21 W. Le montage (figure \ref{MONTAGE}) comprend une alimentation
de courant continu de marque Convergie/Fontaine (type
ASF1000/20.50) permettant une lecture directe de l'intensité du
courant $I$ délivré, et un voltmètre numérique de marque METRIX
avec lequel on mesure la chute de tension $U$ aux bornes de la
lampe.

En faisant varier l'intensité du courant $I$ depuis de très
faibles valeurs (proches de zéro) jusqu'à environ 2.5 A, la mesure
de la chute de tension $U$ aux bornes de la lampe permet de
déterminer la résistance $R$ de la lampe pour chaque valeur de $I$
en appliquant la loi d'Ohm ($R=U/I$). Lors de l'expérience, il est
nécessaire d'attendre que la lampe se stabilise en température
avant de relever le couple de valeurs $(U,I)$.

Un soin tout particulier est nécessaire pour déterminer la
résistance de la lampe à température ambiante $R_0$ qui
conditionne la qualité et la précision de la procédure de
calibration. Les points de mesures $(R, I)$  à très faible
intensité ont été extrapolés à courant nul en utilisant un modèle
parabolique de la forme : $R(I) = R_0 + A I^2$. Une procédure
d'ajustage par la méthode des moindres carrés permet de déterminer
le paramètre $A$ et la valeur de la résistance à température
ambiante $R_0$.

\begin{figure}[h!]
\centerline{\includegraphics[scale=0.9,angle=-90]{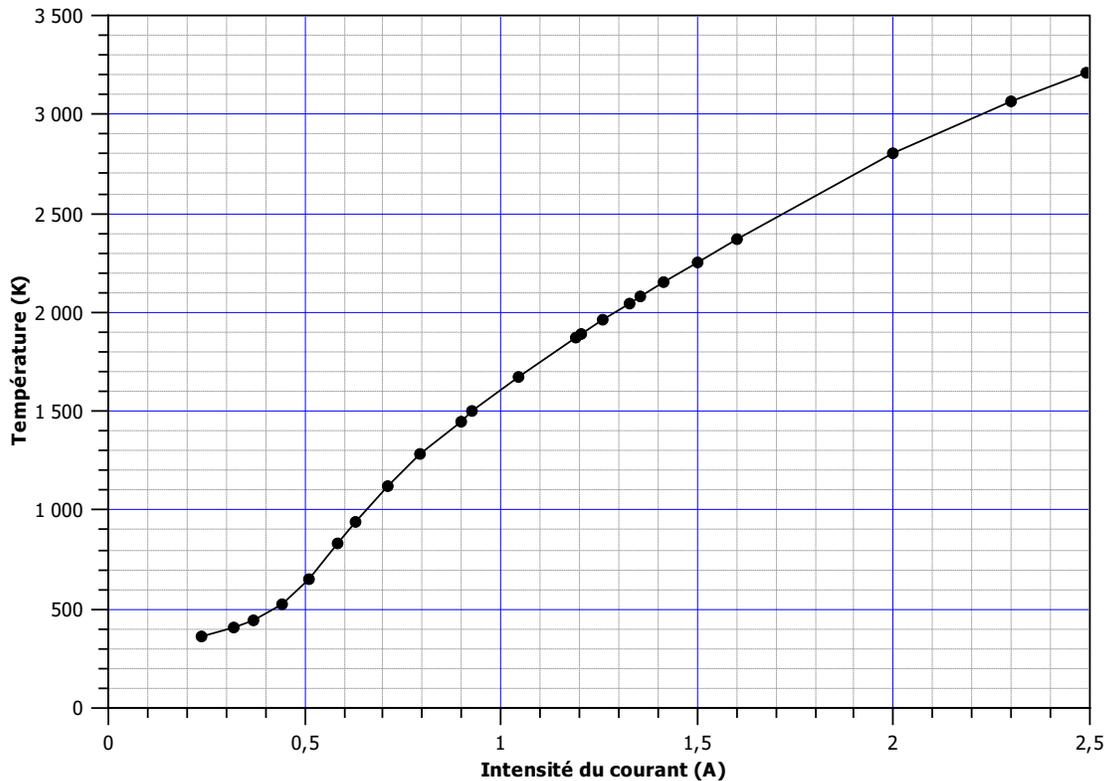}}
\caption{\textit{Courbe de calibration donnant la température du
filament de la lampe en fonction de l'intensité du courant.} }
\label{TfoncdeI}
\end{figure}

Connaissant la valeur expérimentale du rapport $R(T)/R_0$ pour
chaque valeur de l'intensité du courant, on est en mesure de
déterminer la température $T$ du filament en utilisant la relation
(\ref{FIT}) (résolution d'une équation du second degré). Au final,
on obtient une courbe de calibration sous la forme de la
température $T$ du filament en fonction de l'intensité du courant
$I$ (figure \ref{TfoncdeI}).

D'un point de vue purement énergétique, le filament de la lampe
chauffé est le siège de la transformation de l'énergie électrique
sous forme de chaleur et sous forme de rayonnement. En considérant
qu'un état stationnaire est atteint, on a l'égalité entre la
puissance $P=UI$ dissipée par effet Joule, et la somme des
puissances mises en jeu par transfert thermique entre le filament
à température $T$ et le \og milieu ambiant \fg{} à température
$T_a$ et par rayonnement (loi de Stefan-Boltzmann). En appelant
$\alpha$ le coefficient d'échange thermique incluant les
phénomènes de conduction et de convection thermiques,
$\varepsilon$ l'émissivité du tungstène, et $S$ la surface
émissive du filament nous avons :
\begin{equation}
UI = \alpha \left( T- T_a \right) + S \varepsilon \sigma T^4.
\label{EQUATIONUI}
\end{equation}
\`A partir des mesures de $U$ et $I$, il est simple de calculer la
puissance mise en jeu, puis de considérer la quantité $UI/T^4$. En
effet, selon l'équation (\ref{EQUATIONUI}), nous avons :
\begin{equation}
\frac{UI}{T^4} = \alpha \frac{\left( T- T_a \right)}{T^4} + S
\varepsilon \sigma.  \label{EQUATIONUISURT4}
\end{equation}
L'équation (\ref{EQUATIONUISURT4}) comporte deux termes ; le
premier est prépondérant à \og basse température \fg{}, et le
second terme est constant, sous réserve que $\varepsilon$ soit
constant en fonction de la température.

\begin{figure}[h!]
\centerline{\includegraphics[scale=.9,angle=-90]{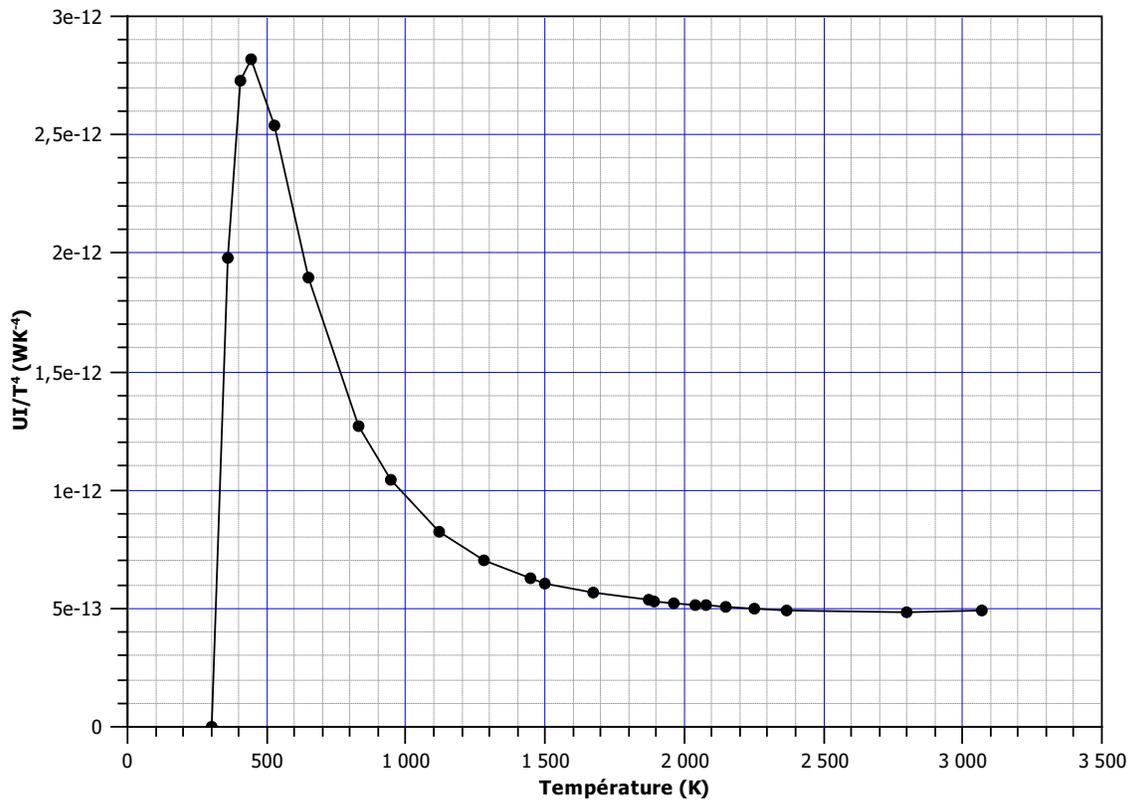}}
\caption{\textit{Courbe donnant le rapport $UI/T^4$ en fonction de
$T$.} } \label{RAPPORTUIsurT4PHILIPS}
\end{figure}

Sur la figure \ref{RAPPORTUIsurT4PHILIPS}, on a représenté le
graphe de la quantité $UI/T^4$ en fonction de $T$, \textit{qui est
tout à fait conforme aux prévisions annoncées plus haut, et qui
prouvent la validité des mesures}. Pour les températures élevées,
la courbe est horizontale et permet de déterminer la surface
émissive du filament de tungstène.

\begin{figure}[h!]
\centerline{\includegraphics[scale=.6]{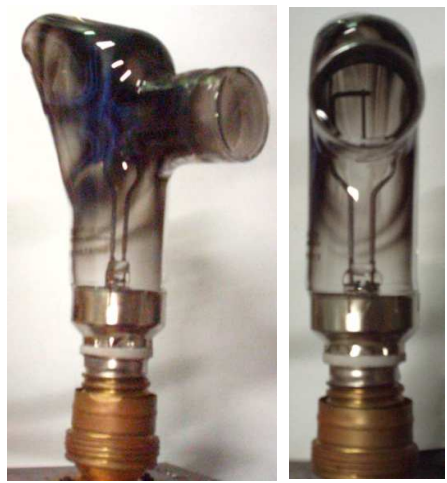}}
\caption{\textit{Photographie de la lampe étalon à ruban de
tungstène.} } \label{photoosram}
\end{figure}

\noindent\textbf{4. Procédure appliquée à une lampe étalon}\\
La procédure de calibration en température présentée plus haut a
été appliquée à une lampe étalon à ruban de tungstène (lampe OSRAM
WI 17G figure \ref{photoosram}) étalonnée sur une valeur unique de
l'intensité du courant $I$. L'étalonnage par le constructeur
indique une température de 2689~K pour une intensité égale à
14.1~A.

\begin{figure}[h!]
\centerline{\includegraphics[scale=.5,angle=-90]{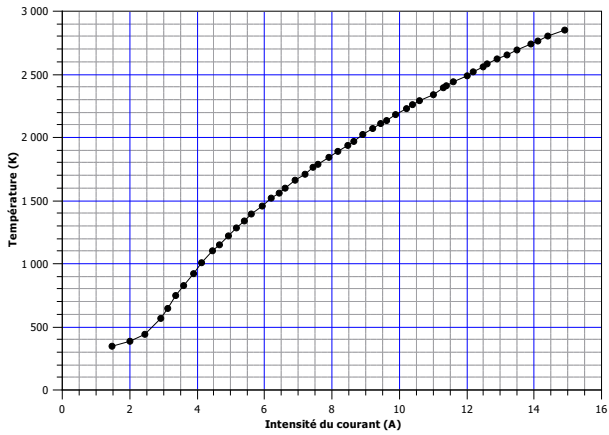}
\includegraphics[scale=.5,angle=-90]{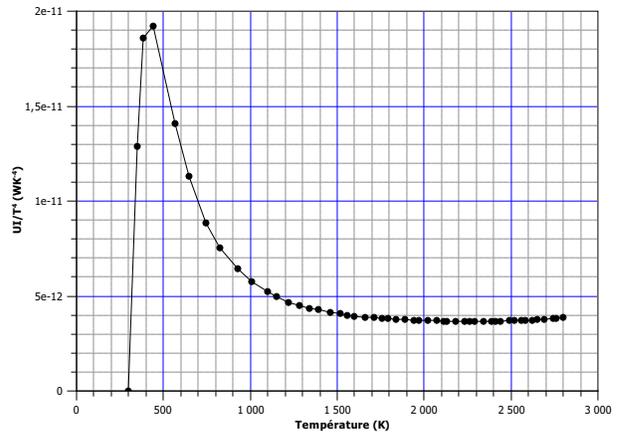}}
\caption{\textit{Courbes donnant la température $T$ de la lampe à
ruban de tungstène en fonction de l'intensité du courant (gauche)
et le rapport $UI/T^4$ en fonction de $T$ (droite).} }
\label{COURBESlampeOSRAM}
\end{figure}

Les courbes donnant la température de la lampe en fonction de
l'intensité du courant et le rapport $UI/T^4$ en fonction de $T$
sont présentées sur la figure \ref{COURBESlampeOSRAM}. La valeur
de la température déterminée pour l'intensité de 14.1~A (intensité
de calibration) est égale à 2760~K, ce qui correspond à une erreur
relative de 2.6\% par rapport à la température de calibration
annoncée. Compte tenu du phénomène de vieillissement de cette
lampe, avec une évaporation du tungstène et son dépôt sur la paroi
interne de l'ampoule visible sur la figure \ref{photoosram}, il
est certain que la calibration n'est plus \og optimale \fg{} et
permet d'expliquer l'écart entre la température mesurée pour
14.1~A et la température de calibration.

\noindent\textbf{5. Conclusion et perspectives}\\
La procédure de calibration d'une lampe à filament de tungstène
présentée dans cette communication est relativement simple à
mettre en {\oe}uvre, et à la portée de tout laboratoire de
recherche et développement. La validation de la procédure
est réalisée grâce à l'évaluation du rapport $UI/T^4$ en fonction
de $T$, dont le comportement est tout à fait conforme à ce que
prévoit la théorie.

Enfin, pour être en mesure de calibrer un capteur photométrique à
partir du rayonnement émis par un filament en tungstène dont on
connaît la température de surface, il est nécessaire de considérer
la relation (\ref{Planck}) donnant la luminance spectrale du corps
noir multipliée par l'émissivité du tungstène fonction de la
température $T$ et de la longueur d'onde $\lambda$ (voir tableau
\ref{TABLO1}).
\begin{table}[h!]
\begin{tabular}{|c|r|r|r|r|r|r|r|r|r|r|}
\hline
           &                                                                                \multicolumn{ 10}{|c|}{Longueur d'onde en $\mu$m} \\
\hline
Température (K) &       0.25 &        0.3 &       0.35 &        0.4 &        0.5 &        0.6 &        0.7 &        0.8 &        0.9 &          1 \\
\hline
      1600 &      0.448 &      0.482 &      0.478 &      0.481 &      0.469 &      0.455 &      0.444 &      0.431 &      0.413 &       0.390 \\
\hline
      1800 &      0.442 &      0.748 &      0.476 &      0.744 &      0.465 &      0.452 &       0.440 &      0.425 &      0.407 &      0.385 \\
\hline
      2000 &      0.436 &      0.474 &      0.473 &      0.747 &      0.462 &      0.448 &      0.436 &      0.419 &      0.401 &      0.381 \\
\hline
      2200 &      0.429 &     0.470 &       0.470 &      0.471 &      0.458 &      0.445 &      0.431 &      0.415 &      0.896 &      0.378 \\
\hline
      2400 &      0.422 &      0.465 &      0.466 &      0.468 &      0.455 &      0.441 &      0.427 &      0.408 &      0.391 &      0.372 \\
\hline
      2600 &      0.418 &      0.461 &      0.464 &      0.464 &      0.451 &      0.437 &      0.423 &      0.404 &      0.386 &      0.369 \\
\hline
      2800 &      0.411 &      0.456 &      0.461 &      0.461 &      0.448 &      0.434 &      0.419 &        0.400 &      0.383 &      0.367 \\
\hline
\end{tabular}
\caption{Emissivité du tungstène en fonction de la température et
de la longueur d'onde [Lide, 1991-1992].\label{TABLO1}}
\end{table}


\end{document}